\documentstyle[prl,aps,twocolumn,epsf]{revtex}
\begin{document}
\draft 
\twocolumn[ 
\hsize\textwidth\columnwidth\hsize\csname@twocolumnfalse\endcsname 

\title{Dynamics of Activated Escape, and Its Observation in a
Semiconductor Laser}

\author{J. Hales$^{(a)}$,
A. Zhukov$^{(b)}$, R. Roy$^{(c)}$, and M.I.~Dykman$^{(b)}$\cite{byline1}} 

\address{$^{(a)}$ CREOL, University of Central Florida,
Orlando, Fl 32816\\
$^{(b)}$Department of Physics and Astronomy,
Michigan State University, East Lansing, Michigan 48824\\
$^{(c)}$Department of Physics, University of Maryland, College
Park, MD 20742}

\date{\today} 
\maketitle
\widetext 
\begin{quote}
We report a direct experimental observation and provide a theory of
the distribution of trajectories along which a fluctuating system
moves over a potential barrier in escape from a metastable state. The
experimental results are obtained for a semiconductor laser with
optical feedback. The distribution of paths displays a distinct peak,
which shows how the escaping system is most likely to move. We argue
that the specific features of this distribution may give an insight
into the nature of dropout events in lasers.
\end{quote}
] 
\narrowtext 


Fluctuation-induced escape plays an important role in many physical
phenomena, from traditionally studied diffusion in solids and protein
folding to
switching in lasers \cite{Willemsen,Yacomotti}, resonantly driven
trapped electrons \cite{Gabrielse}, and systems which display
stochastic resonance \cite{RPP,SR}. In the analysis of escape, it is
important to be able not only to calculate, but also to control the
escape probability. To do this one has to know how the system {\it
moves} when it escapes.

Escape is an example of a large fluctuation.  If fluctuations are
small on average, for most of the time the system wanders near the
initially occupied metastable state $q_a$ and only occasionally moves
far away from it.  The central idea of the theory of large
fluctuations is that paths to a remote state $q_f$ lie within a narrow
tube centered at an {\it optimal} path to this state $q_{\rm
opt}(t|q_f,t_f)$ \cite{Onsager,Grareview}, where $t_f$ is the instant
of reaching $q_f$. Optimal paths reveal determinism of motion in large
fluctuations. They can be observed by analyzing the prehistory
probability density (PPD) $p_h(q,t|q_f,t_f)$ for a system to have
passed through a point $q$ at time $t$ provided the system had been
fluctuating about the stable state for a long time and reached $q_f$
at time $t_f$.  For given $t_f-t>0$, $p_h(q,t|q_f,t_f)$ should peak
for $q$ lying on $q_{\rm opt}(t|q_f,t_f)$ \cite{prehistory}. The
sharply peaked PPDs have indeed been observed, but so far only in
analog and digital simulations \cite{RPP}, and for points $q_f$ lying
inside the attraction basin of $q_a$
\cite{private}.

In the present paper we analyze the dynamics of the system during
escape and, using a semiconductor laser with optical feedback, provide
a direct experimental observation of the prehistory distribution. This
distribution displays a distinct peak, as seen from Fig.~2 below. We
show that such peak arises even for final states lying {\it behind}
the boundary of the domain of attraction to the initially occupied
metastable state, e.g. behind the top of the potential barrier in
Fig.~1.
We reveal qualitative features of the PPD, relate them to escape
dynamics, and compare the theoretical and experimental results.

In the analysis of escape dynamics we will use a simple model of an
one-variable overdamped system which performs Brownian motion in a
metastable potential $U(q)$, with equation of motion

\begin{equation}
\label{Langevin}
\dot q = -U'(q) + \xi(t), \; \langle
\xi(t)\xi(t')\rangle = 2D\delta (t-t').
\end{equation}

\noindent
Here, $\xi(t)$ is zero-mean white Gaussian noise. We assume that the
noise intensity is small compared to the height of the potential
barrier, $D\ll \Delta U$, where $\Delta U=U(q_b)-U(q_a)$, see Fig.~1
($q_a$ and $q_b$ are the positions of the local minimum and maximum of
$U(q)$). In this case the escape rate $W\propto \exp(-\Delta U/D)$ is
small compared to the characteristic reciprocal relaxation time
$t_r^{-1} = U^{\prime\prime}(q_a)$.

\begin{figure}
\begin{center}
\epsfxsize=3.0in                
\leavevmode\epsfbox{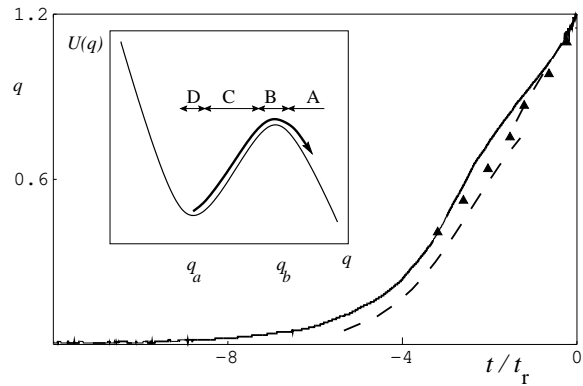}
\vspace{-0.1in}
\end{center}

\caption{The positions of the maxima of the prehistory probability
density with respect to the coordinate $q$ for given time $t$ (solid
line), and with respect to $t$ for given $q$ (triangles). The data of
simulations refer to a Brownian particle (\ref{Langevin}),
(\ref{potential}), the noise intensity $D=1/60$ ($\Delta U/D=10$),
$q_f=1.2$, and $t_r=1$. The dashed line shows the asymptotic results
(\ref{harmonic}), (\ref{inside_2}) for $t_m(q)$. Inset: escape from a
potential well; the motion in the regions A-D is discussed in the
text.}
\end{figure}

On its way to a point $q_f$ behind the barrier, the system is expected
to move differently in the four regions shown in Fig.~1. In the region
A behind the barrier top it should move nearly along the noise-free
trajectory $\dot q = -U'$.
In the region B near the barrier top, where $|q-q_b|\alt l_D$ ($l_D =
(2D/\lambda)^{1/2}$ is the diffusion length, $\lambda =
|U^{\prime\prime}(q_b)|$), the influence of noise becomes
substantial. The motion is diffusive and is controlled by
average-strength fluctuations. The system stays here for the Suzuki
time $ t_{_{\rm S}}=\lambda^{-1}\ln |q_b-q_a|/l_D$ \cite{Suzuki}.
In the region C ($|q-q_b|\gg l_D$) the system is driven by the noise
$\xi(t)$ against the regular force $-U^{\prime}$, which requires a
strong outburst of noise. For Gaussian noise, the probabilities of
different appropriate realizations of $\xi(t)$ differ from each other
exponentially strongly. Therefore there is an optimal realization of
noise, which is much more probable than others. It corresponds to an
optimal path of the system $q_{\rm opt}$. For fluctuations from the
attractor to an intrawell state  such path is given by \cite{Grareview}

\begin{equation}
\label{opt_path}
\dot q_{\rm opt} = U^{\prime}(q_{\rm opt}).
\end{equation}
%
In the region D near the attractor, $|q-q_a|\alt (Dt_r)^{1/2}$, the
system performs small fluctuations before a large fluctuation leading
to escape occurs.

Diffusive motion near the barrier top $q_b$ gives rise to a strong
broadening of the distribution of fluctuational paths. If the
destination point $q_f$ approaches $q_b$ from inside the well, the
distribution width diverges in the bounce-type approximation
\cite{prehistory}. As we show, the divergence
disappears if one goes beyond this approximation. The analytic
solution will be obtained
assuming that $\ln[\Delta U/D]\gg 1$. This condition is {\it not}
needed for the physical picture of the escape dynamics to apply, as we
demonstrate experimentally and through simulations.

For a Markov system (\ref{Langevin}), the PPD can be written as

\begin{eqnarray}
\label{definition}
p_h(q,t|q_f,t_f)={\rho(q_f,t_f|q,t)\rho(q,t,|q_i,t_i)
\over \rho(q_f,t_f|q_i,t_i)}
\end{eqnarray}

\noindent
where $\rho(q_1,t_1|q_2,t_2)$ is the probability density of the
transition from $q_2$ at the instant $t_2$ to $q_1$ at the instant
$t_1$ ($t_1>t_2$). We choose the initial instant $t_i$ so that $
W^{-1} \gg t_f-t_i > t-t_i \gg t_r$. In this time range the system
forgets its the initial intrawell state $q_i$. The statistical
distribution inside and outside the well (not too far from the barrier
top) is quasistationary, $\rho(q,t|q_i,t_i) = \rho(q)$, and can be
easily calculated.


The prehistory distribution $p_h$ has a simple form for $q$ and $q_f$
lying behind the barrier top $q_b$ in the region A in Fig.~1. For
brevity, we give $p_h(q,t|q_f,t_f)$ in the case where $q,q_f$ are both
in the range where $U(q)$ is {\it parabolic} near $q_b$, but $q_f$ is
far enough behind $q_b$, $q_f-q_b\gg l_D$. The transition
probability density $\rho(q_f,t_f|q,t)$ for such $q,q_f$ is known
\cite{Kampen}, and from (\ref{definition})

\begin{equation}
\label{harmonic}
p_h(q,t|q_f,0)= (2 z_f/l_D)r(q)e^{\lambda t}
\exp\left[-(z-z_fe^{\lambda t})^2\right],
\end{equation}

\noindent 
(we have set $t_f=0$). Here, $z \equiv z(q)=[1-\exp(2\lambda
t)]^{-1/2}\tilde q$, $z_f \equiv z(q_f)$ (note that $t< 0$),
$r(q)=\exp(\tilde q^2)\left[1-{\rm erf}(\tilde q)\right]/2$, and
$\tilde q= (q-q_b)/l_D$.

For $|t|\alt \lambda^{-1}$, the distribution (\ref{harmonic}) has a
sharp Gaussian peak as a function of $q$, with width $\propto
l_D$. Behind the barrier, the peak lies on the noise-free trajectory
$\dot q = U^{\prime}(q)=-\lambda (q-q_b)$, which arrives at $q_f$ for
$t=0$.


Interestingly, the PPD peak remains {\it sharp}, with width $\sim l_D$, even
where its maximum reaches the barrier top, which happens for $t=
t(q_b) = -\lambda^{-1}\ln[\pi^{1/2}(q_f-q_b)/l_D]$.

For earlier times $-t>-t(q_b)$,
the system is mostly on the {\it intrawell} side of the barrier, and
for large $|t/t(q_b)|$ the peak of $p_h$ as a function of $q$ moves
away from the harmonic range.  Of interest is the position $t_m(q)$ of
the peak of $p_h$ as a function of time for given $q$. It shows when
the particle was most likely to pass through the point $q$ before
arriving at $q_f$. Inside the well, for $q_b-q \gg l_D$, the time
$t_m$ and the integral width of the PPD $\gamma (q)= 1/p_h(q,t_m)$ are
of the form

\begin{eqnarray}
\label{int_2}
\lambda t_m= -\ln[2(q_b-q)(q_f-q_b)/l_D^2],\; 
\gamma = e|q_b-q|.
\end{eqnarray}

\noindent
From (\ref{int_2}), $t_m$ depends on $q_b-q$ logarithmically. In
contrast, $\gamma(q)$ grows linearly with $q_b-q$. It becomes {\it
parametrically} larger than the distribution width $\gamma\sim l_D
\propto D^{1/2}$ at the barrier top and outside the well.

Far from the barrier top in the region C in Fig.~1, the motion of the
system is determined by large fluctuations against the force
$-U^{\prime}(q)$. 
In this region, $p_h(q,t|q_f,t_f)$ can be obtained using the
Smoluchowski equation which follows from (\ref{definition}),

\begin{equation}
\label{convolution}
p_h(q,t|q_f,t_f)=\int dq'\,p_h(q,t|q',t')\,
p_h(q',t'|q_f,t_f).
\end{equation}

\noindent
It is convenient to choose $t'$ in (\ref{convolution}) so that the
major contribution to the integral over $q'$ came from $q'$ lying on
the internal side of the barrier close to $q_b$ and yet away from the
diffusion region, $q_b-q_a \gg q_b-q' \gg l_D$. Then the second
integrand in (\ref{convolution}) is given by (\ref{harmonic}).  

The distribution $p_h(q,t|q',t')$ as a function of $q'$ can be
obtained from (\ref{definition}) by solving, in the eikonal
approximation, the Fokker-Planck equation for $\rho(q',t'|q,t) =
\exp[-S(q',t'-t|q,0)/D]$. To zeroth order in $D$, $S$ satisfies a
Hamilton-Jacobi equation for the action of an auxiliary dynamical
system \cite{Grareview}. An appropriate Hamiltonian trajectory of this
system gives the optimal path $q_{\rm opt}(t'-t|q,0)$ for a
fluctuation in which the original system (\ref{Langevin}) starts at
the point $q$ and moves further away from the attractor
\cite{q_opt_ref}. This path is given by Eq.~(\ref{opt_path}). The
major contribution to the integral (\ref{convolution}) comes from the
points $q'$ which lie close to this path. For small $\delta q'=
q'-q_{\rm opt}(t'-t|q,0)$, it suffices to keep quadratic in $\delta
q'$ terms in $S$, and then $p_h(q,t|q',t')$ is Gaussian in $\delta
q'$. In the appropriately chosen parameter range,
the time $t'$ drops out from
(\ref{convolution}), and one obtains \cite{to_be_published}

\begin{eqnarray}
\label{inside_2}
&&p_h(q,t|q_f,0) = \left|\lambda/U^{\prime}(q)\right|M\exp(-M),
\end{eqnarray}

\noindent
where $M=M(q,t)= -\lambda(q_f-q_b) \left[q_{\rm
opt}(-t|q,0)-q_b\right]/D$.

Eq.~(\ref{inside_2}) describes the distribution of trajectories along
which the escaping system moves inside the well. This distribution has
a distinct peak. For given $q$, the peak is located for $M(q,t_m(q))=
1$. From (\ref{opt_path}), 
the position of the peak
obeys the equation $dt_m/dq = 1/U'(q)$. This means that, inside the
potential well, the particle is most likely to move along the optimal
path (\ref{opt_path}). In a multi-dimensional system, the peak of
$p_h$ will lie on the most probable escape path, which goes from the
attractor to the saddle point. 

The distribution (\ref{inside_2}) is strongly asymmetric, both in $q$
and $t$. The integral width

\begin{equation}
\label{int_3}
\gamma(q) = 1/p_h(q,t_m|q_f,0) =e|U^{\prime}(q)|/\lambda
\end{equation}

\noindent
is {\it independent} of the noise intensity and is {\it nonmonotonic}
as a function of $q$. It is maximal for $U^{\prime\prime}(q)=0$ where
the velocity along the optimal path is maximal. 
The broadening of the tube of escape paths in time comes largely from
the area near the barrier top. However, it is ``amplified'' as it is
carried away by the trajectories flow, and therefore it is maximal where
the flow is most fast.

\begin{figure}
\begin{center}
\epsfxsize=2.9in                
\leavevmode\epsfbox{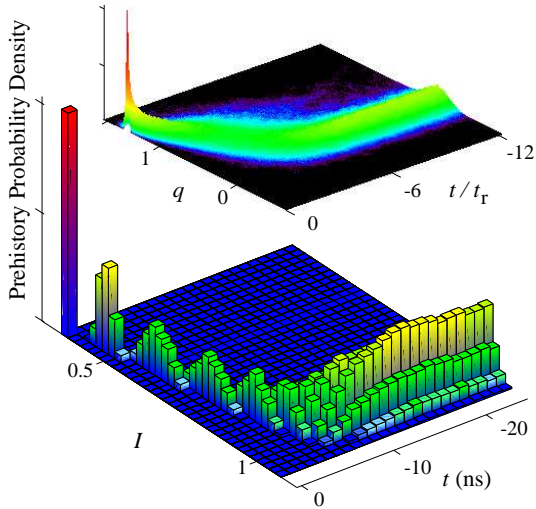}
\vspace{-0.1in}
\end{center}
\caption{(color) The prehistory probability distribution of the scaled
radiation intensity $I$ for experimentally observed dropout events in
a semiconductor laser.  Inset: the PPD for a Brownian particle,
obtained from simulations for the same parameters as in Fig.~1.}
\end{figure}


As $-t$ increases further, the peak of the distribution
(\ref{inside_2}) approaches the diffusion region $|q-q_a|\sim
(Dt_r)^{1/2}$ near the potential minimum, and the peak width
(\ref{int_3}) again shrinks down. For large $-t$, the PPD
(\ref{definition}) goes over into the stationary distribution
$\rho(q)$, which has a nearly Gaussian peak at $q_a$ with variance
$Dt_r/2$. 

We note that, in the most interesting region C, the positions of the
maxima of $p_h$ (\ref{inside_2}) with respect to $q$ for given $t$ and
with respect to $t$ for given $q$ are different. This indicates that
there is no well-defined most probable escape path in space and time,
which would go from the metastable state all the way over the barrier
top. Still one can tell when the escaped system passed, most
probably, through a given point, and where the system was most
probably located at a given time.

The discussed qualitative features of the prehistory distribution can
be seen from the results of digital simulations for the model
potential

\begin{equation}
\label{potential}
U(q)=q^2/2 - q^3/3,
\end{equation}

\noindent
The data were obtained using a standard algorithm \cite{Mannella}, and
refer to  8000 events.

A distinct peak of the simulated $p_h$ is seen in Fig.~2. The peak of
$p_h$ as a function of $q$ changes with increasing $|t|$ from a narrow
Gaussian near $q_f$ to a broad and asymmetric between $q_a$ and $q_b$,
and then again to a comparatively narrow Gaussian near $q_a$.  The
positions of the peak of $p_h$ with respect to time, $t_m(q)$, and
coordinate, $q_m(t)$, are compared in Fig.~1. Both curves are close to
each other. Outside the well they practically coincide and closely
follow the noise-free path of the system, $dt_m/dq = -1/U'(q)$. The
motion displays a characteristic slowing down near the barrier top.
Inside the well the peak moves close to the optimal fluctuational
path $\dot q = U'(q)$.
The distribution $p_h$ becomes time-independent for large $|t|$. Therefore
$t_m(q)$ is well-defined only for $q$ not too close to the potential
minimum $q_a$.

The data of simulations in Figs.~1, 2 refer to the noise intensities
$D/\Delta U=0.1$, where the asymptotic analytical theory applies only
qualitatively. In particular, the expressions (\ref{harmonic}) and
(\ref{inside_2}) for $p_h$ in different ranges of $q$ do not merge
with each other smoothly, as seen from Fig.~1. However, there is good
qualitative agreement between the analytical and numerical data,
including the position of the peak and the integral width of the PPD.

Numerical results on the standard deviation of the PPD $\sigma$ for
two noise intensities are shown in Fig.~3. 
As expected, the distribution width reaches its maximum well inside
the well, near the inflection point $U''(q)=0$. For higher $D$, the
maximum is less pronounced.

\begin{figure}
\begin{center}
\epsfxsize=2.6in                
\leavevmode\epsfbox{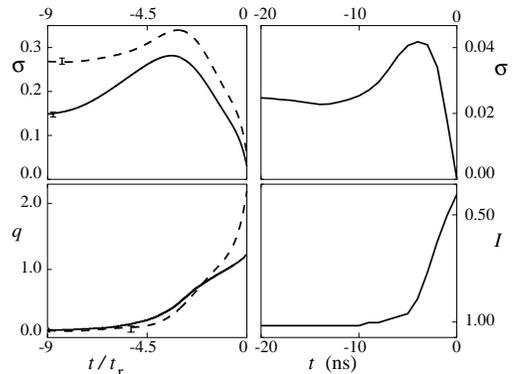}
\vspace{-0.1in}
\end{center}
\caption{Standard deviation and the position of the maximum of the
prehistory distribution at given time, for simulated Brownian motion
(left panel) and a semiconductor laser with optical feedback (right
panel). Solid and dashed lines on the left panel refer to $\Delta
U/D=10$ and 3, respectively. The scales in the panels are different
(see the text).}
\end{figure}


The experimental observation of the prehistory distribution was made
using a semiconductor laser with optical feedback. The setup was
similar to that used before \cite{Roy_Hohl} and consisted of a
temperature-stabilized laser diode and a remote flat-surface
mirror. The feedback could be controlled by a variable attenuator
between them. Near the solitary laser threshold, such a system is
unstable: after some time of nearly steady operation the radiation
intensity drops down, then it comparatively fast recovers to the
original value, then drops down again, etc. In the experiment, the
intensity output was digitized, with time resolution 1~ns. To obtain
the prehistory distribution, the intensity records were superimposed
backward in time, starting from the instant at which the intensity, on
its way down, reached a certain level (10\% above the extreme dropout
point). The PPD obtained from 1512 events for the feedback 15.63\%
is shown in Fig.~2.

The mechanism of power dropouts is vividly discussed in the literature
\cite{Roy_Hohl,Yacomotti,Huyet,Vaschenko,Hohl}. Most authors agree
that the role of noise in this effect is crucial.  A simple model
\cite{Henry} describes the dropouts in terms of activation escape of
the light intensity $I$ over a potential barrier with shape
(\ref{potential}). Previous observations \cite{Roy_Hohl} were in
agreement with this model, which motivated us to measure the
prehistory distribution for dropout events.

It is seen from Figs.~2 and 3 that 
the results of the observations agree with major qualitative results
on noise-induced escape. The experimental PPD displays a distinct
peak. The shape of this peak is similar to the shape of the PPD of a
noise-driven system, with the light intensity $I$ playing the role of
the coordinate $q$ ($I$ is scaled by its metastable value). The peak
is narrow at small time $|t|$, and displays a characteristic
broadening at intermediate times.  For larger $|t|$, the peak becomes
time-independent.  From the data in Figs.~2 and 3, the relaxation time
of the system is $t_r\approx 2$~ns. From the value of the escape rate
$W\approx 5\times 10^{-3}$~ns$^{-1}$ found in \cite{Roy_Hohl}, it
follows that, for the model ~(\ref{potential}), $\Delta U/D \approx
3$. Using an estimate $\sigma_0\approx
(D/U^{\prime\prime}(q_a))^{1/2}$ for $\sigma$ at large $-t$, one can
estimate the difference $6(\Delta U/D)^{1/2}\sigma_0$ in the light
intensity $I$ at the minimum and maximum of the potential
(\ref{potential}). It then follows from Fig.~3b that the system goes
through the potential maximum for $t\sim -4$~ns, i.e. the width
$\sigma$ reaches its maximum near the potential maximum. In
combination with larger $\sigma_{\max}/\sigma_0$ compared to that in
Fig.~3a for the same $\Delta U/D$, this indicates that the model
(\ref{Langevin}), (\ref{potential}) is oversimplified. However, the
overall form of the PPD seen from the data provides an important
argument in favor of the stochastic model of dropout events. We expect
that it will be possible to use high-resolution data on the prehistory
distribution in order to establish a quantitative model of the system.

In conclusion, we have analyzed the dynamics of a system in activated
escape and revealed its distinctive features related to the occurrence
of optimal paths and to the motion slowing down near a barrier
top. The escape trajectories lie within a well-defined tube, and the
system is most likely to go through a cross-section of this tube at a
well-defined time before it is found behind the barrier. For the first
time, a tube of escape trajectories has been observed in experiment,
by analyzing dropout events in a semiconductor laser.

The work at MSU was partly supported from the NSF funded MRSEC and the
NSF grants no. PHY-9722057 and no. DMR-9809688.




\end{document}